\documentstyle[aps,psfig,epsf,multicol]{revtex}
\begin{document} 
\draft
\title{Disorder-enhanced delocalization and 
local-moment quenching in a disordered antiferromagnet} 
\author{Avinash Singh,\cite{asingh} Martin Ulmke, and Dieter Vollhardt} 
\address{Theoretische Physik III, 
Elektronische Korrelationen und Magnetismus,
Universit\"{a}t Augsburg, D-86135 Augsburg, Germany}
\maketitle
\begin{abstract} 
The interplay of disorder and spin-fluctuation effects 
in a disordered antiferromagnet is studied.
In the weak-disorder regime ($W\le U$), 
while the energy gap decreases rapidly with disorder,
the sublattice magnetization, including quantum corrections,
is found to remain essentially unchanged
in the strong correlation limit.
Magnon energies and N\'{e}el temperature are {\em enhanced} by 
disorder in this limit.
A single paradigm of disorder-enhanced delocalization qualitatively 
accounts for all these weak disorder effects.
Vertex corrections and magnon damping, which appear only at order $(W/U)^4$,
are also studied.
With increasing disorder a crossover is found at $W\sim U$, 
characterized by a rapid decrease in sublattice magnetization due to 
quenching of local moments, and formation of spin vacancies.
The latter suggests a spin-dilution behavior which
is indeed observed in 
softened magnon modes, lowering of N\'{e}el temperature, and 
enhanced transverse spin fluctuations.
\end{abstract} 
\begin{multicols}{2}\narrowtext
\section{Introduction}
The manifestation of quantum antiferromagnetism in parent cuprates,\cite{vaknin}
discovered soon after the birth of high-$T_{\rm c}$ superconductivity in
the doped materials,\cite{bednorz} has led to intensive efforts to understand
the nature of this phase both within the Hubbard model, 
as well as its strong-coupling counterpart, the quantum Heisenberg model.\cite{brenig}  
Features such as the
substantially reduced  sublattice magnetization 
(relative to the classical value) as 
deduced from neutron-scattering experiments,\cite{vaknin}
the substantially enhanced linewidth observed in the two-magnon Raman
scattering experiments,\cite{lyons,rrpsingh}
as well as a detailed fitting of the temperature dependence
of the spin correlation length\cite{chn,cs}  have confirmed the importance of
quantum spin fluctuations in these low-dimensional, low-spin systems.
Antiferromagnetic (AF) spin correlations are also present in 
other strongly correlated systems, notably the transition-metal oxides
such as NiO, $\rm V_2 O_3$, $\rm LaVO_3$, $\rm NiS_{2-x}Se_x$, 
and heavy-fermion compounds such as 
YbP, $\rm U_2 Zn_{17}$, $\rm UCd_{11}$, $\rm UCu_5 $
which exhibit AF ordering of d and f electrons,
respectively, in their ground states.\cite{stewart,anderson,edwards,ott} 

Many of the correlated electron systems are intrinsically disordered, 
and the metal-insulator transition observed in several amorphous materials
such as doped semiconductors, 
amorphous ${\rm Ge_{1-x}Au_x}$ and ${\rm B_{1-x}Cu_x}$ alloys, and granular alumina
have the character of both the Mott transition and the
Anderson localization transition.\cite{lee}
The role of strong disorder effects has also been emphasized in 
the recently-studied transition-metal oxides such as 
${\rm LaNi_{1-x}Co_x O_3}$ and ${\rm Na_xWO_3}$.\cite{akray1}
The square root dip in the electronic density of states near the Fermi energy,
characteristic of disorder-induced enhancement of interaction effects,\cite{lee}
has been found to change to a linear form and then to a soft quadratic gap
on the insulating side.\cite{akray2,akray1}
In some cases, such as in $\rm La_{1-x}Sr_x VO_3$, $\rm V_2 O_3$,
the insulator-metal transition is accompanied with loss of 
AF order,\cite{akray1,edwards}
whereas $\rm NiS_{2-x}Se_x$ exhibits an AF metallic phase.\cite{edwards}

It is therefore of interest to study 
the interplay of quantum spin fluctuation and disorder effects.
Of particular interest are questions such as: 
(i) are transverse spin fluctuations and quantum corrections to the
sublattice magnetization enhanced by disorder? 
(ii) Does a gapless AF state exist, and, if so, at what critical disorder 
strength is AF long-range order (AFLRO) destroyed?
(iii) Is the AF state destabilized by disorder at finite temperature, 
such that the N\'{e}el temperature (for dimensions $d>2$) is lowered?  
To answer these questions we will examine the influence of diagonal disorder
on various properties of the Hubbard antiferromagnet, such as
sublattice magnetization, quantum spin fluctuations, 
magnon energy and damping, N\'{e}el temperature, Hubbard energy gap,
and electronic density of states. 

Recently spin fluctuation effects were examined in
impurity-doped antiferromagnets, both within the 
Heisenberg model,\cite{bulut,kampf,manousakis,dkumar,poilblanc}
as well as the Hubbard model,\cite{gapstate,impscatt,magimp} 
in order to study magnetic dynamics 
in cuprate antiferromagnets doped with nonmagnetic and magnetic
impurities such as Zn, Al, Ga, and Fe, Ni, Co 
respectively.
It was found that a static vacancy, created by the 
replacement of a fermion with a nonmagnetic impurity, for instance,   
leads to strong magnon scattering.
It is therefore also of interest to contrast scattering
of magnons caused by disorder with that by static vacancies.
The third, related, case 
is that of magnon scattering off mobile vacancies, 
as in hole-doped cuprates, which is of course much more efficient at destroying
AFLRO; spin correlation lengths of order of $1/\sqrt{x}$
for hole concentration $x$ have been reported from 
neutron-scattering studies.\cite{spincor}

This study therefore complements earlier works on the disordered 
Hubbard model where other aspects have been studied,  such as
the metal-insulator transition,\cite{mit,belitz} local-moment formation,\cite{msb,lw,sachdev} 
phase diagram etc. A variety of methods have been used earlier, including
the scaling theory,\cite{mcmillan} field-theoretic 
approaches,\cite{mafrad,finkelstein,cast}
renormalization group (RG),\cite{fink2,castellani} real-space RG,\cite{ma,yi}
slave-boson formulation,\cite{zim} dynamical mean field theory,\cite{juv,dk,ulmke} 
quantum Monte Carlo studies,\cite{ulmke2} and unrestricted 
Hartree-Fock theory together with random phase approximation,
and Onsager-reaction-field correction to mean-field 
theory of equivalent spin models.\cite{logan,logan2,logan3}
The disordered Hubbard model exhibits extremely rich physics
and contains the non-interacting Anderson localization transition, the 
purely interacting magnetic transition, and of course the non-trivial fixed
point describing the metal-insulator transition in the disordered, 
interacting theory. In addition, various ingredients such as the phenomena
of weak localization, disorder-induced enhancement of interaction effects, 
leading to singularities at the Fermi energy, local-moment behavior, etc.,
are also contained.\cite{belitz}
Generally, the simultaneous presence of interaction and disorder leads to a 
new coupling of the quantum degrees of freedom in two-particle quantities
that has no counterpart in non-interacting, disordered or interacting, 
pure systems.

We consider the following Hubbard Hamiltonian with
random on-site energies, and with a filling of one fermion per site, so
that an AF ground state is obtained, 
\begin{equation} 
H = \sum_{i\sigma} \epsilon_i \hat{n}_{i\sigma}
 -t\sum_{\langle ij \rangle \sigma}
(\hat{a}_{i \sigma}^{\dagger} \hat{a}_{j \sigma} + {\rm h.c.}) +
U\sum_{i} \hat{n}_{i \uparrow} \hat{n}_{i \downarrow}  .
\end{equation} 
The random on-site energies $\epsilon_i$ are chosen from a uniform 
distribution with $-W/2<\epsilon_i <W/2$, 
the distribution width $W$ parametrizing the disorder 
strength. We consider both the strong correlation
limit, with the correlation term $U$ much larger than the free-particle
bandwidth $B=2Zt$, where $Z=2d$ is the coordination number,
as well as the intermediate correlation regime, with $U\sim B$.  
For concreteness, we consider the square lattice, generalization to
three dimension and to other bipartite lattices being straightforward.

We will use several methods/approximations in this paper.
The broken-symmetry state is obtained in the 
unrestricted Hartree-Fock approximation (UHF), and 
transverse spin fluctuations about this state 
are studied in the Random Phase Approximation (RPA).
Disorder is treated both perturbatively as well as within
a numerical diagonalization approach on finite lattices.
In the latter approach the eigenfunctions and eigenvalues
of the HF Hamiltonian in the fully self-consistent state 
are used to obtain sublattice magnetization,
energy gap, and the magnon spectrum. 
The T-matrix approach used earlier for impurities,\cite{gapstate} 
is also employed for comparison. 
Quantum spin-fluctuation corrections 
are obtained at the one-loop level.\cite{quantum}
The N\'{e}el temperature, energy gap, and electronic spectrum 
are also studied within the dynamical mean field theory (DMFT).\cite{ulmke}

The outline of the paper is as follows.
Section II deals with the reduction of the Hubbard gap
due to formation of disorder-induced states. 
Disorder renormalizations of the magnon energy, 
damping, and sublattice magnetization 
are described in sections III and IV, based on results of a
perturbative analysis in powers of $W/U$, 
discussed in the Appendix.
A qualitative explanation of the disorder effects 
is given in Section V in terms
of the notion of disorder-enhanced delocalization.
Section VI describes the crossover to the
strong disorder regime $(W>U)$, 
where the electronic spectrum is
gapless, and spin vacancies are
formed due to quenching of local moments. 
Magnon softening, enhancement in transverse spin fluctuation 
due to spin vacancies, and robustness of the 
gapless AF state are discussed. 
Conclusions are given in section VII. 

\section{Disorder-induced states in the gap}
In this section we examine the formation of disorder-induced states 
within the Hubbard gap, which reduce the 
charge gap, and determine the critical disorder strength at which 
the gap vanishes. We use the T-matrix approach and a numerical UHF
approach, described in subsections A and B respectively.
The T-matrix approach is exact for a single impurity,
and has been used earlier to study 
the formation of defect states within the gap
due to a single nonmagnetic impurity in the Hubbard AF.\cite{gapstate} 
In order to use this approach for the disordered AF, 
with random potentials on {\em every site},
we make a {\em local approximation}
in which we consider a single site, and treat the 
random potential $\epsilon_i$ on this site as an impurity potential.
Comparison with results of the 
numerical UHF analysis, in which disorder is treated exactly,
indicates that this approximation 
actually works quite well, particularly in the strong correlation limit,
where states are strongly localized. 
Spin fluctuation processes will lead to small changes in the energy gap of order $J$ 
in the strong correlation limit.

\subsection{T-matrix approach}
Within this approach the location of impurity-induced states
is obtained from the pole in the T-matrix,
$T_{\sigma}(\omega)=
\epsilon_i/(1-\epsilon_i[g^{0}_{\sigma}(\omega)]_{ii})$,
where $[g^{0}_{\sigma}]_{ii}$ is the local host Green's function.
For the pure AF it is given (in the HFA) by,
$[g^{0}_{\sigma}(\omega)]_{ii}
=(1/N)\sum_{\bf k} (\omega\mp \sigma \Delta)/
(\omega^2 - E_{\bf k} ^2)$,
for site $i$ in the A or B sublattice.\cite{gapstate}
Here $2\Delta=mU$ is the Hubbard energy gap in the pure AF,
$E_{\bf k}=\sqrt{\Delta^2 +\epsilon_{\bf k}^2}$ is the AF band energy, 
and $\Delta$ is obtained from the self-consistency condition
$(1/N)\sum_{\bf k} (2E_{\bf k})^{-1} = U^{-1}$.

Now, for a given disorder strength $W$,
the inverse potential (absolute value) 
has a lower bound $1/|\epsilon_i | > 2/W$. 
Therefore poles in the T-matrix are present for
$|[g^{0}_{\sigma}]_{ii}|>2/W$, so that
disorder-induced states are formed within the gap, 
as shown by hatched regions in Fig. 1.
If $-\tilde{\Delta}$ and $\tilde{\Delta}$
mark the energies (shown by arrows) up to which
\begin{figure}
\vspace*{-70mm}
\hspace*{-28mm}
\psfig{file=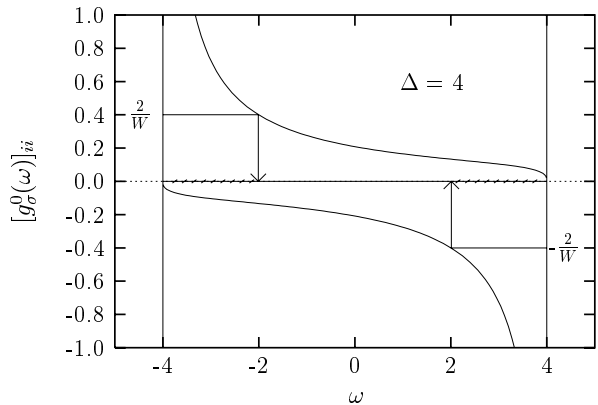,width=135mm,angle=0}
\vspace{-70mm}
\caption{Local host Green's function
$[g^0_{\sigma}(\omega)]_{ii}$ vs. $\omega$ in the pure AF state.
Intersections with lines $2/W$ and $-2/W$
show the extent to which disorder-induced states are formed within
the Hubbard gap. In all figures energies are in units of the hopping
parameter $t$.}
\end{figure}
\noindent
states are formed,
then $\tilde{\Delta}$ is obtained from 
$[g^{0}_{\sigma}(-\tilde{\Delta})]_{ii}= 2/W $.
The remaining Hubbard gap $2\tilde{\Delta}$ is 
thus obtained from the solution of
\begin{equation}
\frac{1}{N}\sum_{\bf k} \frac{\Delta + \tilde{\Delta} }{E_{\bf k}^2 -\tilde{\Delta}^2 }
=\frac{2}{W} .
\end{equation}
A plot of the normalized energy gap 
$2\tilde{\Delta}/2\Delta$ is shown in Fig. 2 as a function of the relative
disorder strength $W/U$ for $U/t=10$,
indicating an almost linear reduction with disorder strength. 

With increasing disorder strength states are formed deeper 
in the Hubbard gap, and 
when states have approached from both sides 
in the middle of the Hubbard gap, the energy gap vanishes.
The critical disorder strength $W_c$ at which the energy gap 
$2\tilde{\Delta}$ vanishes is therefore given by 
$[g_{\sigma}^{0}(0)]_{ii}=2/W_c $, 
which yields the following equation for the
critical disorder strength,
\begin{equation}
\frac{1}{N}\sum_{{\bf k}}\frac{\Delta}{\Delta^2 + \epsilon_{\bf k} ^2} = \frac{2}{W_c}  .
\end{equation}

Considering the strong correlation limit as a special case, 
and keeping terms up to order $t^2/U^2$, 
the critical disorder strength is then given by $W_c /U =1+8t^2/U^2 $,
where we used $m=1-8t^2/U^2$ and $(1/N)\sum_{\bf k} \epsilon_{\bf k}^2 =4t^2$ 
for the square lattice.
Thus, with decreasing interaction strength the ratio $W_c /U$ actually 
{\it increases}.
This is because the kinetic energy  becomes relatively important with
decreasing interaction strength, and the bandwidth starts competing with
disorder strength. 

\subsection{Numerical UHF analysis}
In order to check the validity of the T-matrix approximation,
we have also used a numerical UHF analysis.
In this approach, the HF Hamiltonian on a finite lattice is
\begin{figure}
\vspace*{-70mm}
\hspace*{-28mm}
\psfig{file=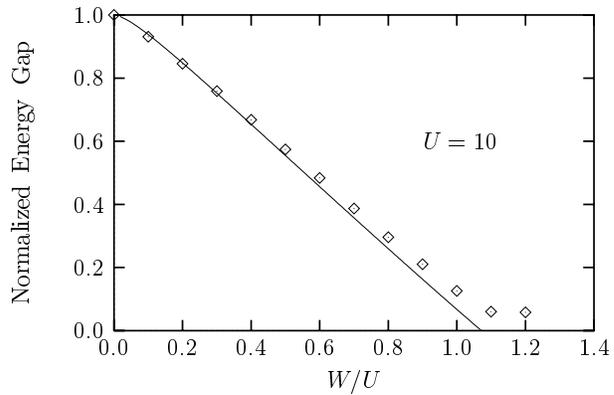,width=135mm,angle=0}
\vspace{-70mm}
\caption{Normalized energy gap vs. $W/U$ at $T=0$, from
the T-matrix analysis (line), and the numerical UHF analysis 
for a $10\times 10$ lattice (squares).} 
\end{figure}
\ \\
\noindent
numerically diagonalized self-consistently, 
so that disorder is treated exactly.
This approach has been described 
earlier in the context of hole/impurity doping
in the Hubbard antiferromagnet.\cite{uhf1,uhf2}
The energy gap is obtained from the 
energy difference between the lowest energy state of the
upper Hubbard band and the highest energy state of the lower Hubbard band.
Configuration averaging is performed over 100 different 
realizations of the random on-site potentials
on a $10\times 10$ lattice.
The reduction in energy gap 
with disorder strength using this method is also shown in Fig. 2.
The numerical analysis shows a saturation of the
energy difference at $W/U\sim 1$ due to finite system size. 
Deviations from the T-matrix approach 
are more pronounced at lower interaction strengths where 
the fermion states are more extended. 
The almost linear reduction of energy gap with disorder is also
seen at finite temperatures, as shown in Fig. 3. The critical disorder
strength decreases rapidly with increasing temperature.  

\section{Magnon renormalization}
Magnons are the low-energy excitations associated with transverse
spin fluctuations in the broken-symmetry state of systems possessing
continuous spin-rotational symmetry. Therefore they play an important
role in several macroscopic properties 
such as the temperature-dependence of the order parameter,
N\'{e}el temperature, specific heat, etc. 
We therefore consider the magnon propagator and obtain 
the disorder-induced renormalizations in magnon energies,
transverse spin correlations, and the quantum spin-fluctuation
correction to sublattice magnetization. 
The magnon propagator with site indices $i,j$ is defined in terms of
spin-lowering  and spin-raising operators by $G^{-+}_{ij}=
\langle \Psi_{\rm G} | T [ S^- _i (t) S^+ _j(t')]|\Psi_{\rm G}\rangle$.
We take the Green's function approach and write the RPA result 
in $\omega$-space as
\begin{figure}
\vspace*{-70mm}
\hspace*{-28mm}
\psfig{file=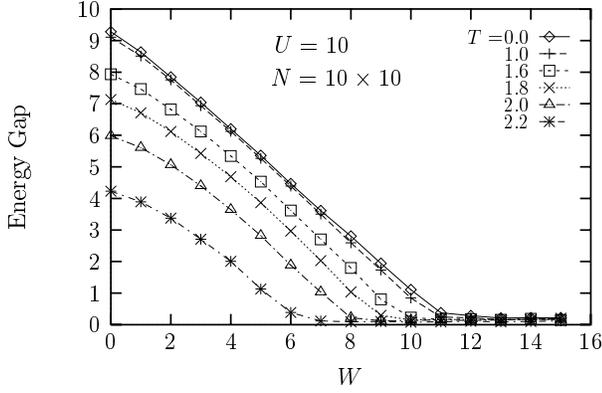,width=135mm,angle=0}
\vspace{-70mm}
\caption{Energy gap vs. $W$ from the numerical UHF analysis
at different temperatures (shown against plotting symbols).}
\end{figure}
\noindent
\begin{equation}
G^{-+}=
\frac{\chi^0_{\rm dis}}{1-U\chi^0_{\rm dis}}
= \frac{G^0}{1-\Sigma G^0}  .
\end{equation}
Here the matrix (in site indices) $\chi^0_{\rm dis}$
is the zeroth-order antiparallel-spin particle-hole propagator
for the disordered AF,
with matrix elements given by
$[\chi^0_{\rm dis}(\omega)]_{ij} =i\int(d\omega'/2\pi)
[g_{\uparrow}(\omega')]_{ij}
[g_{\downarrow}(\omega'-\omega)]_{ji}$,
written in terms of the one-particle Green's function
$[g_\sigma (\omega)]_{ij}$, and evaluated 
in the self-consistent, broken-symmetry state.
By $\chi^0_{\rm pure}$ we define the corresponding 
quantity for the pure AF, in terms of which the matrix 
$G^0=\chi^0_{\rm pure}/(1-U\chi^0_{\rm pure})$ 
is the magnon propagator for the pure AF.
Furthermore, the disorder self energy $\Sigma = U^2 \delta\chi^0$
is expressed in terms of the disorder-induced perturbation
$\delta \chi^0 \equiv \chi^0_{\rm dis} - \chi^0_{\rm pure}$.
Details of the perturbative analysis for $\delta \chi^0$
in the strong coupling limit
are given in the Appendix, and we use the result here 
for the disorder self energy $\Sigma$ 
which has diagonal and nearest-neighbor (NN) terms.
For the pure AF in the strong coupling limit and in the 
two-sublattice basis (indices A,B), the propagator in
${\bf k}$-space takes the form,\cite{uhf1} 
\begin{equation}
[G^0]^{-1}({\bf k},\omega)=
\left [
\begin{array}{lr}
1+\omega & \gamma_{\bf k} \\
\gamma_{\bf k} & 1-\omega
\end{array} \right ],
\end{equation}
in energy units where $2J=8t^2/U=1$. 
Here $\gamma_{\bf k}\equiv (\cos k_x +\cos k_y)/2$,
so that in real space 
$[G^0]^{-1}$ has only diagonal elements,
$[G^0]^{-1} _{ii}=1\pm \omega$ for site $i$ in A and 
B sublattices, and NN matrix elements
$[G^0]^{-1} _{i,i+\delta}= 1/Z$,
where $i+\delta$ refers to NN of $i$.

We first consider the configuration-averaged self energy 
$\langle \Sigma \rangle=U^2 \langle \delta\chi^0 \rangle$.
This approximation neglects vertex corrections which, however, appear only at
order $(W/U)^4$, and are discussed separately in the next  section. 
From the results given in the Appendix 
for the matrix elements of $\delta \chi^0$ 
up to order $(W/U)^2$, we obtain
$\langle \Sigma_{ii}\rangle =-\sigma$
and $\langle \Sigma_{i,i+\delta}\rangle = -\sigma/Z $, 
in units of $U^2t^2/\Delta^3$, where $\sigma \equiv (1/6)W^2/U^2 $.  
Substituting the translationally symmetric
$\langle \Sigma \rangle$ in Eq. (4), Fourier transforming 
$[G^0]^{-1} - \langle \Sigma \rangle$, and finally inverting we obtain
the following expression for the renormalized magnon propagator in
${\bf k}$-space and the two-sublattice basis, 
\begin{eqnarray}
G^{-+}({\bf k},\omega)&=&
-\frac{1}{2}\frac{1}{\sqrt{1-\gamma_{\bf k} ^2}}
\left [ \begin{array}{lr} 
1-\frac{\omega}{2\tilde{J}} & -\gamma_{\bf k} \\
-\gamma_{\bf k} & 1+\frac{\omega}{2\tilde{J}} \end{array}\right ]\nonumber \\
&\times &
\left ( \frac{1}{\omega-\tilde{\omega}_{\bf k} +i\eta}
-\frac{1}{\omega+\tilde{\omega}_{\bf k}-i\eta}
\right ) ,
\end{eqnarray}
containing both the retarded and advanced parts having poles 
below and above the real-$\omega$ axis, respectively.
Here $\tilde{J}\equiv J(1+\sigma)$, 
and $\tilde{\omega}_{\bf k}=2\tilde{J}\sqrt{1-\gamma_{\bf k} ^2}$
is the disorder-renormalized magnon energy.
Thus at ${\cal O}(W/U)^2$ the form of the magnon propagator
remains unchanged, and there is only a momentum-independent multiplicative
renormalization of magnon energies, leading to an effective
stiffening of the magnon modes by disorder.  
An upward shift of the magnon band 
in the strong-correlation and weak-disorder limit has 
also been observed in a numerical 
RPA study in three dimensions.\cite{logan2}

This effective enhancement of the magnon
energy scale can  also be viewed as resulting from  the enhancement in the 
configuration-averaged NN exchange energy 
$\langle t^2/(U+\epsilon_i -\epsilon_j)\rangle $
which, to second order in $W/U$, is $(1+\sigma)t^2/U$.  
To the extent
that the finite-temperature reduction in sublattice magnetization due to
thermal excitation of magnons is suppressed by this enhancement,
the disordered AF exhibits a {\em slower} $m(T)$ vs.  $T$ falloff, and
therefore a {\em higher} N\'{e}el temperature (for $D>2$). 
For strong coupling,
the N\'{e}el state is therefore {\it stabilized} by weak disorder, 
as also reported in the DMFT\cite{ulmke}
and the Onsager-reaction-field\cite{logan3} studies.

\subsection*{Quantum correction to sublattice magnetization}
As only the effective exchange energy scale gets modified in Eq. (6),
while the form is not changed by disorder,\cite{uhf1} the magnitude
of equal-time, same-site
transverse spin correlations $\langle S_i^- S_i^+ \rangle $
and $\langle S_i^+ S_i^- \rangle $ remain unchanged.
These transverse spin correlations are obtained by
frequency integration of  
the diagonal elements of transverse spin propagators,
$\langle S^- _i (t) S^+ _i (t')\rangle
=-i \int (d\omega/2\pi)
[G^{-+}(\omega)]_{ii} \exp\{-iw(t-t')\}$,
where the appropriate part (retarded or advanced) 
of $G^{-+}$ is taken 
depending on whether $t'<t$ or $t'>t$. 
%
%
Taking $t'\rightarrow t^-$,
and using the retarded part of $G^{-+}$ from Eq. (6), we obtain
\begin{eqnarray}
\langle S_i^- S_i^+ \rangle &=&
\frac{1}{2}\frac{1}{N}
\sum_{\bf k} \left [\frac{1}{\sqrt{1-\gamma_{\bf k}^2}} -1 \right ] \nonumber \\
\langle S_i^+ S_i^- \rangle  &=&
\frac{1}{2}\frac{1}{N}
\sum_{\bf k} \left [\frac{1}{\sqrt{1-\gamma_{\bf k}^2}} +1 \right ] .
\end{eqnarray}
Here the second result is obtained using the relationship
$G^{+-}_{AA}(\omega)=G^{-+}_{BB}(\omega)$ 
which follows from spin-sublattice symmetry.
The result for total transverse spin fluctuation 
$\langle S^- _i S^+ _i + S^+ _i S^- _i \rangle$, is thus 
identical to the RPA result for the pure Hubbard AF,\cite{quantum}
as well as the spin-wave-theory (SWT) result for the 
Quantum Heisenberg Antiferromagnet (QHAF).\cite{anderson.swt,oguchi}
Therefore up to order $(W/U)^2$ the quantum spin-fluctuation
correction to sublattice magnetization  in the strong coupling limit 
remains unchanged from the spin-wave-theory result $\delta m_{\rm SF}=
(1/N)\sum_{\bf k} [(1-\gamma_{\bf k}^2)^{-1/2} -1] 
\approx 0.39$ in two dimensions.\cite{anderson.swt,oguchi}

\section{Vertex Corrections and Magnon Damping}
In this section we consider vertex corrections which were neglected
in the previous section due to use of the configuration-averaged
self energy $\langle \Sigma\rangle $ in Eq. (4).
As at higher orders this approximation produces terms like 
$G^0\langle{\Sigma}\rangle G^0\langle{\Sigma}\rangle G^0$ in Eq. (4), 
we therefore subtract out this term and focus here on
the configuration-averaged proper self energy at second order,
\begin{equation}
\Gamma=\langle \Sigma G^0 \Sigma\rangle -\langle \Sigma\rangle 
G^0\langle \Sigma\rangle  ,
\end{equation}
which precisely incorporates the vertex corrections. 
As shown later in this section, this second-order scattering
process results in magnon damping, which therefore only appears
at order $(W/U)^4$. 
We note here that the term ``magnon damping''
in this section refers to the decay of a momentum mode,
and not to the decay into particle-hole excitations,
which is energetically ruled out in the strong-correlation limit
where $J<<U$.

Considering the matrix element $\Gamma_{ij}$, we note that since 
$\Sigma$ is only limited to diagonal and NN matrix elements,
if sites $i$ and $j$ are far apart, then there are no correlations between the
two $\Sigma$ terms, and the difference vanishes. Therefore the proper
self energy $\Gamma$ arises only from local correlations in the disorder
self energy terms, and has diagonal, 
NN and next-nearest-neighbor (NNN) matrix elements.
The vertex corrections therefore result in new NNN spin couplings in the magnon
propagator, besides renormalizing the NN couplings. 
Such longer-range spin couplings also arise in the Hubbard 
AF at intermediate and weak couplings.

For the square lattice ($Z=4$), we give below the results
for matrix elements of $\Gamma$ for the case $i$ in A sublattice ($i\in A$);
results for the other case following from symmetry. 
Also given are the results after substitution of the matrix
elements of $G^0({\bf k},\omega)$, obtained from Eq. (5).
We now illustrate the evaluation of $\Gamma$ for the 
diagonal matrix element. Expanding the matrix product, 
and using the property of the disorder self energy that 
the diagonal element $\Sigma_{ii}$ equals the 
sum of the NN elements $\Sigma_{i,i+\delta}$,
this can be written as 
$\Gamma_{ii}=[
(\langle \Sigma_{i,i+\delta}\Sigma_{i+\delta', i}\rangle
-\langle \Sigma_{i,i+\delta}\rangle\langle\Sigma_{i+\delta', i}\rangle)
 \times \\
(G^0_{ii}+G^0_{i+\delta, i}+G^0_{i,i+\delta'}+G^0_{i+\delta,i+\delta'})] 
$, where summation over $\delta$ and $\delta'$ is implied.
Configuration averaging, with
$\sigma_2\equiv \langle\epsilon_i ^2\rangle /U^2$
and $\sigma_4\equiv \langle \epsilon_i ^4\rangle /U^4$,
for the second and fourth moments, 
yields
$(\langle \Sigma_{i,i+\delta}\Sigma_{i+\delta', i}\rangle
-\langle \Sigma_{i,i+\delta}\rangle\langle\Sigma_{i+\delta', i}\rangle)
=Z^{-2}[(\sigma_4-\sigma_2^2) + (\sigma_4+3\sigma_2^2)\delta_{\delta\delta'}]$.
Substituting $G^0_{ij}(\omega)=
(1/N)\sum_{\bf k} G^0({\bf k},\omega)\exp\{i{\bf k}.({\bf r}_i -{\bf r}_j)\}$,
and taking the appropriate matrix elements of $G^0({\bf k},\omega)$ in the
two-sublattice basis, depending on sublattices of sites $i$ and $j$, 
yields the expression for $\Gamma_{ii}$ in Eq. (9).
Similarly evaluating the NN and NNN elements,
with $i+\delta$, $i+\kappa$, and $i+\kappa'$ 
standing for the NN, NNN (diagonal), and NNN (straight)
of $i$ respectively, we obtain,
\end{multicols}
\widetext
\begin{eqnarray}
\nonumber \\
\Gamma_{ii}&=&(\sigma_4-\sigma_2 ^2)
[G^0_{AA}+(G^0 _{AB}+G^0 _{BA})\gamma_{\bf k}
+G^0 _{BB}\gamma_{\bf k} ^2] 
+ Z^{-1}(\sigma_4 +3\sigma_2 ^2)
[G^0_{AA}+(G^0 _{AB}+G^0 _{BA})\gamma_{\bf k}
+G^0 _{BB}] \nonumber \\
&=&(\sigma_4-\sigma_2 ^2)
[(1-\omega)(1-\gamma_{\bf k}^2)/(\omega_{\bf k}^2-\omega^2)]
+ Z^{-1}(\sigma_4 +3\sigma_2 ^2)
[2(1-\gamma_{\bf k}^2)/(\omega_{\bf k}^2 -\omega^2)] \nonumber \\ 
\nonumber \\
\Gamma_{i,i+\delta}&=& Z^{-1} (\sigma_4-\sigma_2 ^2)
[(G^0 _{AA}+G^0 _{BB})(1+\gamma_{\bf k} ^2)
+ 2(G^0 _{AB}+G^0 _{BA})\gamma_{\bf k} ] 
+4Z^{-2}\sigma_2 ^2 [(G^0 _{AA}+G^0 _{BB}) +
(G^0_{AB}+G^0_{BA})\gamma_{\bf k}] \nonumber \\
&=& Z^{-1} (\sigma_4-\sigma_2 ^2)[2(1-\gamma_{\bf k}^2)/
(\omega_{\bf k}^2-\omega^2)] 
+ 4Z^{-2}\sigma_2 ^2 [2(1-\gamma_{\bf k}^2)/(\omega_{\bf k}^2-\omega^2)] 
\nonumber \\ 
\nonumber \\
\Gamma_{i,i+\kappa}&=&Z^{-2}
(\sigma_4-\sigma_2 ^2)2[G^0 _{AA} \cos k_x \cos k_y 
+ (G^0_{AB}+G^0_{BA})\gamma_{\bf k}
+G^0_{BB}] \nonumber \\
&=&Z^{-2}(\sigma_4-\sigma_2 ^2)2[\{2(1-\gamma_{\bf k}^2)-(1-\omega)
(1-\cos k_x\cos k_y)\}/
(\omega_{\bf k}^2-\omega^2)] \nonumber \\ 
\nonumber \\
\Gamma_{i,i+\kappa'}&=&Z^{-2}
(\sigma_4-\sigma_2 ^2)[G^0 _{AA} \gamma_{2{\bf k}} +(G^0_{AB}+G^0_{BA})\gamma_{\bf k}
+G^0_{BB}] \nonumber \\
&=&Z^{-2}(\sigma_4-\sigma_2 ^2)[\{2(1-\gamma_{\bf k}^2)-(1-\omega)(1-\gamma_{2{\bf k}})\}/
(\omega_{\bf k}^2-\omega^2)] 
\end{eqnarray}
\ \\
Here the summation over momentum ${\bf k}$ is implied.
A straightforward check confirms that the sum of matrix elements 
$\Gamma_{ii}+Z\Gamma_{i,i+\kappa}+Z\Gamma_{i,i+\kappa'}$ involving sites of the 
same sublattice (diagonal and NNN) exactly equals the sum
$Z\Gamma_{i,i+\delta}$ involving sites on opposite sublattices.
This ensures that the Goldstone mode, which has amplitudes
1 and $-1$ on the two sublattice sites, is preserved, as expected
from the continuous spin-rotational symmetry. 
We also notice that the various terms involve $k$-sums 
of the type $\sum k^2/(c^2 k^2-\omega^2)$ 
from long-wavelength internal magnon modes.
Therefore the self-energy terms are all non-singular in two dimensions.

We now proceed with the magnon renormalization 
due to this proper self-energy correction $\Gamma$
up to order $W^4$. To this order, it is sufficient to examine  
the lowest-order correction $\langle {\bf q} |\Gamma|{\bf q}\rangle$
to the eigenvalue of the $U(1-U\chi^0_{\rm pure})$ 
matrix for the pure AF.
The relevant eigenvalue is $1-\sqrt{\omega^2 + \gamma_{\bf q} ^2}$ 
in energy units such that $2J=1$. 
The magnon amplitudes for state $|{\bf q}\rangle$ 
are $\sin \theta/2$ and $-\cos \theta/2$
in the two-sublattice basis, where 
$\cos\theta=\omega/\sqrt{\omega^2 + \gamma_{\bf q} ^2 }$ and $\sin\theta=
\gamma_{\bf q}/\sqrt{\omega^2 +\gamma_{\bf q} ^2 }$.\cite{impscatt} 
For $\omega=\omega_{\bf q}=\sqrt{1-\gamma_{\bf q}^2}$ 
the magnon energy, these amplitudes
become $\sqrt{1-\omega_{\bf q}}$ and $-\sqrt{1+\omega_{\bf q}} $ respectively. 
Using the matrix elements of $\Gamma$ from above we obtain
for the eigenvalue correction 
$\delta\lambda_{\bf q} ^{(2)}=\langle {\bf q} |\Gamma|{\bf q}\rangle$,
\ \\
\ \\
\begin{eqnarray}
\delta\lambda_{\bf q} ^{(2)}&=&\frac{2}{N}\sum_{i\in A} [
\sin^2 (\theta/2) (\Gamma_{ii} +\Gamma_{i,i+\kappa}Z\cos q_x\cos q_y
+\Gamma_{i,i+\kappa'}Z\gamma_{2{\bf q}})
-\sin(\theta/2) \cos(\theta/2) \Gamma_{i,i+\delta}Z\gamma_{\bf q} ]
\nonumber  \\
&+&  \frac{2}{N}\sum_{j\in B} [
\cos^2 (\theta/2) (\Gamma_{jj} +\Gamma_{j,j+\kappa}Z\cos q_x\cos q_y
+\Gamma_{j,j+\kappa'}Z\gamma_{2{\bf q}})
-\cos(\theta/2) \sin(\theta/2) \Gamma_{j,j+\delta}Z\gamma_{\bf q} ].
\end{eqnarray}
\ \\
In the second term above
(for sites $j$ in the B sublattice)
the matrix elements of $\Gamma$ follow from Eq. (9) 
with $\omega$ replaced by $-\omega$, in view of Eq. (5). 
The magnon energy for mode ${\bf q}$ is then given by the solution of
$1-\sqrt{w^2 +\gamma_{\bf q} ^2} -\delta\lambda_{\bf q} ^{(1)} 
-\delta\lambda_{\bf q} ^{(2)}(\omega) =0$.
Here $\delta \lambda_{\bf q} ^{(1)}=-\sigma(1-\gamma_{\bf q}^2)$ is the 
eigenvalue correction due to the first-order self energy 
$\langle \Sigma\rangle$; its effect on 
magnon stiffening has been discussed earlier. 
\ \\
\ \\

We first consider the magnon renormalization in the long-wavelength ($q \ll  1$),
low-energy ($\omega\ll 1$) limit for simplicity.
We can drop $\omega$ in the numerators in Eq. (9) for the self energy $\Gamma$, 
which removes the sublattice dependence, and the above eigenvalue 
correction simplifies in this limit to
\begin{eqnarray}
\delta\lambda_{\bf q} ^{(2)}(\omega\ll 1)&=&
[\Gamma_{ii} +\Gamma_{i,i+\kappa}Z\cos q_x\cos q_y
+\Gamma_{i,i+\kappa'}Z\gamma_{2{\bf q}}- 
\Gamma_{i,i+\delta}Z\gamma_{\bf q}\sin\theta ]\\
&=&Z[\Gamma_{i,i+\delta}(1-\gamma_{\bf q}\sin\theta) -\Gamma_{i,i+\kappa}(1-\cos q_x\cos q_y)
-\Gamma_{i,i+\kappa'}(1-\gamma_{2{\bf q}})] \nonumber \\
&\approx & \alpha q^2  \nonumber 
\end{eqnarray}
As expected the eigenvalue correction goes like $q^2$. 
Here the identity
$\Gamma_{ii}=Z[\Gamma_{i,i+\delta}-\Gamma_{i,i+\kappa}-\Gamma_{i,i+\kappa'}]$ 
which ensures the preservation of the Goldstone mode has
been used, and the coefficient $\alpha$ is given by
\begin{equation}
\alpha=Z \left ( \frac{\Gamma_{i,i+\delta}}{4} - \frac{\Gamma_{i,i+\kappa}}{2}
-\Gamma_{i,i+\kappa'} \right ) .
\end{equation}
\ \\

Considering now the case of a general $\omega$, 
we find that the
following terms are present in addition to those given in Eq. (11),
\begin{eqnarray}
\delta\lambda_{\bf q} ^{(2)}-
\delta\lambda_{\bf q} ^{(2)}(\omega \ll 1)&=& \omega \cos \theta \;
(1/N)\sum_{\bf k}
[(\sigma_4 -\sigma_2 ^2)(1-\gamma_{\bf k}^2)/
(\omega_{\bf k}^2-\omega^2) \nonumber \\
&-& 2Z^{-1}(\sigma_4-\sigma_2 ^2)\cos q_x\cos q_y(1-\cos k_x\cos k_y)/
(\omega_{\bf k}^2-\omega^2) \nonumber \\
&-& Z^{-1}(\sigma_4 -\sigma_2 ^2)\gamma_{2{\bf q}}(1-\gamma_{2{\bf k}})/
(\omega_{\bf k}^2-\omega^2) ] .
\end{eqnarray}
\begin{multicols}{2}\narrowtext

We now focus on the imaginary part of this second-order correction
$\delta\lambda_{\bf q} ^{(2)}$.
For this purpose we examine the internal momentum sums of the type
$(1/N)\sum_{\bf k} 2\omega_{\bf k} ^2/(\omega_{\bf k}^2-\omega^2)$ which appear in the 
eigenvalue correction above.
Using the following identity for the imaginary part 
\begin{eqnarray}
\Im \frac{1}{N}\sum_{\bf k} 
\frac{2\omega_{\bf k} ^2}{\omega_{\bf k}^2 -\omega ^2}
&=&\pi \frac{1}{N}\sum_{\bf k} \omega_{\bf k} 
\delta(\omega_{\bf k} -\omega) \nonumber \\
&=&\pi \omega N(\omega) ,
\end{eqnarray}
for positive $\omega$, we obtain the imaginary part in terms of 
the magnon density of states $N(\omega)$. We consider the two limiting cases
of low-energy magnon modes $\omega \ll  1$ and high-energy modes with 
$\omega \sim 1 $. For long-wavelength, 
\begin{figure}
\vspace*{-70mm}
\hspace*{-28mm}
\psfig{file=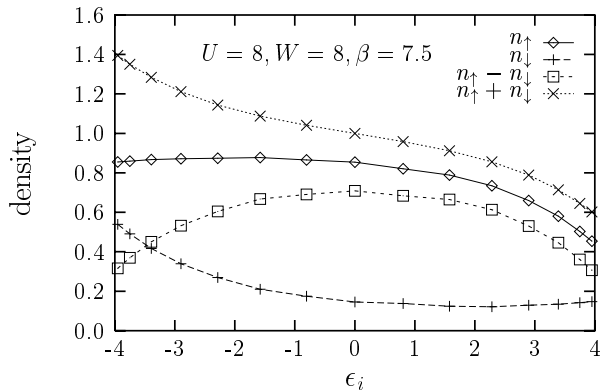,width=135mm,angle=0}
\vspace{-70mm}
\caption{Spin-dependent electronic densities on the A sublattice 
vs. on-site energy in the low-temperature 
ordered state as obtained within the DMFT. 
The band width is chosen $B=8$, as in $d=2$.}
\end{figure}
\noindent
low-energy modes,
$N(\omega)\sim \omega$, therefore 
$\Im \delta \lambda_{\bf q} ^{(2)} \sim Dq^2\omega^2$, where $D=\Im \alpha /\omega^2$ in terms 
of the $q^2$ coefficient given in Eq. (12). 
We thus find the magnon energy to be given by,
\begin{equation}
\omega_{\bf q}=cq[1+iDq^2]
\end{equation}
where $c$ is the renormalized magnon velocity.
The  ratio $\Gamma_{\bf q}/\omega_{\bf q}$ of the magnon damping term to the energy thus vanishes like
$q^2$ in the long-wavelength limit, indicating weak disorder scattering,
caused by the averaging-out of the on-site potential disorder at 
long length scales.

Long-wavelength magnon modes therefore continue to be well-defined 
excitations even with disorder. 
However, for short-wavelength, high-energy modes with energy $\omega \sim 1$,
the presence in the imaginary term of the magnon density of states, which actually
diverges (logarithmically in two dimensions) at the upper band edge at energy $2J$,
indicates that high-energy modes are strongly damped. 
A self-consistent evaluation is therefore required,
with an imaginary term in the internal magnon mode propagator.
This leads to a self-consistent magnon damping of order  $(W/U)^4$. 
In as much as the two-magnon Raman scattering process probes short-wavelength,
high energy magnon modes, this magnon damping of order $(W/U)^4$ will
be important in an analysis of Raman linewidth in disordered antiferromagnets.

\section{Disorder-enhanced delocalization}
All the disorder effects in the AF state obtained so far 
can be understood within a single paradigm of 
{\em disorder-enhanced delocalization}, i.e. an enhancement of the effective
$t/U$ ratio due to disorder. 
Thus, the disorder-induced 
enhancement of magnon energy scale and N\'{e}el temperature 
in strong coupling, 
as well as the reduction of sublattice magnetization (discussed in the Appendix),
can be viewed as arising
from this enhancement of the 
\begin{figure}
\vspace*{-70mm}
\hspace*{-28mm}
\psfig{file=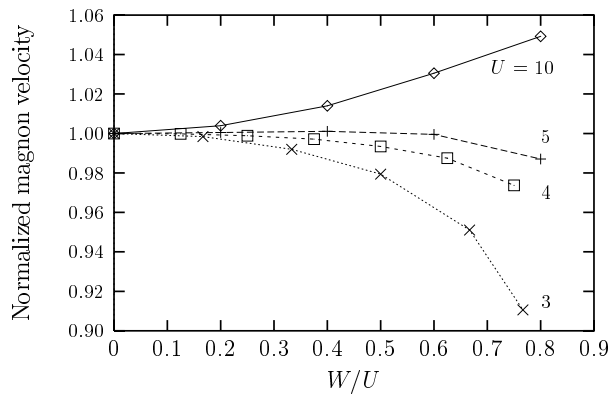,width=135mm,angle=0}
\vspace{-70mm}
\caption{Normalized magnon velocity $c(W)/c(0)$ vs. $W/U$ 
for several interaction strengths, 
obtained from the low-energy magnon modes on a $10\times 10$ lattice.}
\end{figure}
\noindent
effective $t/U$ ratio. 
This delocalization effect of disorder, contrary to its usual tendency to localize,
is characteristic of the AF state
with its inherent localization due to Coulomb barriers,
and can be understood as follows. 
A positive on-site energy $\epsilon_i$ reduces the potential barrier 
for the majority-spin electron, which enhances its probability for tunneling through,
thereby lowering the on-site density. On the other hand, a negative on-site energy 
increases the potential barrier, which has a {\em weaker} effect if the electron is already 
localized. Thus the effects of positive and negative on-site energies 
are {\em asymmetrical}, 
leading to enhanced delocalization on the average. 
This asymmetrical effect is clearly seen in Fig. 4 showing the electronic densities
for different on-site energies, obtained within the DMFT.

This disorder-enhanced delocalization also qualitatively accounts for
{\em opposite} disorder effects on $T_{\rm N}$ 
which are observed for strong and weak coupling.
While for strong coupling, disorder enhances $T_{\rm N}$ 
and stabilizes the N\'{e}el state,
the behavior is reversed for weak coupling.\cite{ulmke,logan3} 
Both behaviors can be understood in terms of an effective $U/t$ ratio
which decreases with disorder.
This is because in the pure Hubbard 
antiferromagnet  $T_{\rm N}$ vs. $U/t$ actually goes through
a maximum, so that with decreasing $U/t$, 
$T_{\rm N}$ is either enhanced or suppressed 
depending on whether one is on the strong- or weak-coupling side of the peak.

Similar effects are also seen
in the magnon velocity, $c=\omega_{\bf q}/q$ in the
limit $q\rightarrow 0$, 
which also shows a peak structure.\cite{uhf1} 
For this purpose we have considered the low-energy magnon modes
on a $10\times 10$ lattice. Magnon energies in the RPA are 
obtained by exact diagonalization of the $\chi^0(\omega)$ matrix,
evaluated in the self-consistent state using the 
numerical UHF approach.\cite{uhf1}
The low-energy modes are four-fold degenerate for the pure AF on a square
lattice, and therefore the normalized magnon velocity is obtained
by averaging $\omega_n(W)/\omega_n(0)$ over the four lowest energy modes. 
The normalized
magnon velocity, configuration-averaged over 20 different configurations,
is shown in Fig. 5, for several values of  $U/t$.
For strong coupling, the magnon velocity increases with disorder,
almost quadratically, in agreement with the perturbative 
result from Eq. (6). The behavior is reversed for weak coupling,
and the magnon velocity decreases with disorder strength.  
As the temperature-dependence of the sublattice magnetization 
in the low-temperature limit depends only on the energy scale
of low-energy magnon modes, this also implies 
slower (faster) temperature-falloff of $m$ 
in the strong (weak) coupling limit.

\section{Strong disorder $(W>U)$ and crossover behavior}
We have seen earlier that 
with increasing disorder strength, 
the energy gap decreases almost
linearly with $W$, and eventually closes when $W\sim U$.
Therefore for $W>U$, assuming a single-occupancy constraint,
the two Hubbard bands would overlap,
indicating that single occupancy for all sites 
is no longer energetically favorable. 
Electrons from the highest-energy sites 
(with $\epsilon_i > U/2$)
are transferred to the lowest-energy sites 
(with $\epsilon_i < -U/2$), 
making them doubly occupied 
and shifting up the energy by $U$. 
The electronic states associated with these 
essentially empty and doubly occupied sites 
are therefore located near the Fermi energy at $U/2$.
As these unoccupied and doubly occupied sites are nonmagnetic, 
the sublattice magnetization starts falling relatively rapidly with 
disorder strength. 
In the limit $t\ll U,W$ the bands are nearly flat 
with bandwidth $\sim W$, and the overlap region is 
$\sim (W-U)$, so that 
a simple estimate for the fraction of these nonmagnetic sites yields 
\begin{equation}
x\sim \frac{W-U}{W} \; , 
\end{equation}
indicating an almost linear decrease in sublattice magnetization
for $W\sim U$.
Thus a crossover takes place at $W\sim U$ from the essentially flat sublattice magnetization
to an almost linear falloff with disorder.
This is clearly seen in Fig. 6 where the configuration-averaged
sublattice magnetization obtained within the UHF approximation
is plotted against the disorder strength $W$. 

However, a more significant consequence of the formation of these nonmagnetic 
sites is that they essentially act like {\it spin vacancies} 
in the antiferromagnet, which leads to spin-dilution behavior,
as discussed in the following subsection. 
The above picture suggests that for $W>U$ the system can be viewed as
a composite of a disordered AF with $\tilde{W}=U$, 
and a spin-diluted system
with a concentration $x=(W-U)/W$ of spin vacancies. 
Strong magnon scattering off static vacancies, leading to
substantial softening of low-energy, long-wavelength modes
and magnon damping has been obtained earlier.\cite{kampf,dkumar,impscatt}
The ratio of magnon damping term
to its energy now goes like
$(W-U)/W$, 
to be contrasted with the small damping ratio
$(W/U)^4 q^2 $ in the weak-disorder regime,
obtained earlier in section IV. 
Vacancy-induced enhancement in transverse spin fluctuation
has not been studied, and we discuss this in the following subsection.

\subsection{Enhanced fluctuations due to spin vacancies}
We present a simple estimate of the enhancement of transverse
spin fluctuations due to spin vacancies. 
We consider the following Hubbard Hamiltonian on a square lattice
with binary-distributed, random NN hopping,
\begin{equation}
H=-\sum_{<ij>\sigma} t_{ij}(\hat{a}_{i\sigma}^{\dagger} \hat{a}_{j\sigma} +{\rm h.c.})
+ U\sum_i \hat{n}_{i\uparrow}\hat{n}_{i\downarrow} \; ,
\end{equation}
where the hopping term $t_{ij}=0$ if sites $i$ or $j$ are vacancy sites,
and $t_{ij}=t$ otherwise. Thus for a vacancy on site $i$,
all hopping terms $t_{i,i+\delta}$ connecting $i$ to its NN sites $i+\delta$ are set to zero.
The vacancy site is thus completely decoupled from the system.
Half-filling is retained by having one fermion per remaining site.
We consider the $U/t\rightarrow\infty$ limit,
where the local moments are fully saturated, and 
the model maps to the localized-spin Heisenberg model.
In this limit the vacancy
problem becomes identical to the spin vacancy problem in the QHAF,
for which magnon renormalization was studied earlier.\cite{kampf}

The structure of the $\chi^0(\omega)$ matrix in the host AF, and 
the modification introduced by spin vacancies has been considered earlier 
in the context of static impurities.\cite{impscatt}
Since the vacancy spin is completely decoupled from the system,
the magnitude of the diagonal matrix element $[\chi^0]_{ii}$ on the
vacancy site $i$ is irrelevant. 
To minimize the perturbation, we treat the vacancy site
as occupied with an isolated spin.  
For a vacancy on site $i$, 
a perturbation is induced in the neighborhood due to the absence of
hopping between vacancy site and NN sites $i+\delta$.
In terms of the notation used in section III, the
following self energy correction $\Sigma=U^2 \delta \chi^0$ is 
obtained
\begin{eqnarray}
\Sigma_{ii} &=& 1 \nonumber \\
 \Sigma_{i,i+\delta} &=& \Sigma_{i+\delta,i}=\Sigma_{i+\delta,i+\delta} =  1/4 
\end{eqnarray}
where again $2J$ has been set to 1. 
Thus, a diagonal contribution $\Sigma _{ii}$ arises if  
a vacancy exists either on site $i$, or on any of the four NN sites $i+\delta$. 
For a finite vacancy concentration $x$,
the probability that a vacancy exists on a site is $x$.
Therefore, configuration-averaging yields 
$\langle \Sigma_{ii}\rangle = 2x$, and 
$\langle \Sigma_{i,i+\delta}\rangle = 2x/4$. 
From Eq. (4) we obtain the same 
expression for the magnon propagator as in Eq. (6), 
except that now $\tilde{J}\equiv J(1- 2x)$, so that 
$\tilde{\omega}_{\bf k}=2J(1-2x)\sqrt{1-\gamma_{\bf k} ^2}$ yields a softening
of the magnon mode, reflecting the spin-dilution behavior.\cite{note1} 
This magnon softening is in contrast to the stiffening 
in the weak-disorder regime, obtained earlier in section III. 
Enhanced thermal excitation of magnons due to this softening
will result in a faster temperature falloff of the sublattice magnetization,
and hence to a lowering of the N\'{e}el temperature.
\begin{figure}
\vspace*{-70mm}
\hspace*{-28mm}
\psfig{file=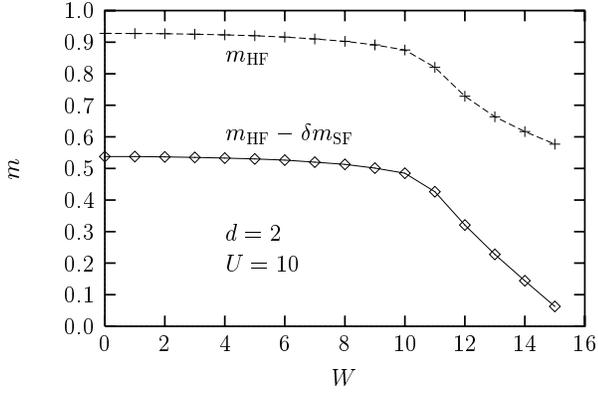,width=135mm,angle=0}
\vspace{-70mm}
\caption{Sublattice magnetization $m$ vs. $W$,
at the HF level for a $10\times 10$ lattice (dashed line),
and including spin fluctuations (solid line).}
\end{figure}

Since the form of the magnon propagator is not changed,
as already discussed in section III, the transverse spin correlations
$\langle S_i^- S_i^+ \rangle $
and $\langle S_i^+ S_i^- \rangle $, as well as 
the lattice-averaged spin-fluctuation correction to 
sublattice magnetization remain unchanged.
Therefore, from the SWT result given after Eq. (7),
$\delta m_{\rm SF} \approx 0.39$ in two dimensions. 
However, as the number of spins is reduced to $1-x$, 
the quantum correction {\em per spin} is {\em enhanced}.

In a recent numerical, finite-size study of the spin-vacancy problem,\cite{trans} 
focussing on the exact evaluation of transverse spin fluctuations in the RPA,
the $x$-dependence of the quantum correction per spin
$\delta m_{\rm SF}/(1-x)$
was found to be best described
by the expression $0.39+0.42x +6.5x^3$.
This results in a nearly vanishing ${\cal O}(x)$ term in $\delta m_{\rm SF}$,
in agreement with the perturbative result obtained above. 
The lattice-averaged sublattice magnetization for the 
disordered AF, 
obtained by accounting for the quantum spin-fluctuation
correction using $m=m_{\rm HF} - \delta m_{\rm SF}$,
with $\delta m_{\rm SF}=(1-x)(0.39+0.42x+6.5x^3)$, 
where the vacancy concentration $x=(W-U)/W$ for $W>U$,
is compared with the HF result in Fig. 6.
The crossover at $W\sim U$ is again clearly seen 
from the rapid decrease in sublattice magnetization.

\subsection{Gapless antiferromagnetic state}
The above result indicates that substantial AF ordering 
remains even for $W\sim U$, where the energy gap closes.
Since AF order persists for $W>U$ 
the gapless AF state is apparently quite robust even in two dimensions.
This feature was also 
observed in the QMC study in $d=2$,\cite{ulmke2}
where substantial AF correlations were seen for $W=U=4t$, while 
the compressibility indicated an absence of the charge gap.
This leads to the possibility, in three dimensions, of a metallic
AF state, if states at Fermi energy are not localized.
Gapless AF states, both metallic and insulating,
were also obtained for
the three-dimensional disordered Hubbard model,
the phase 
\begin{figure}
\vspace*{-76mm}
\hspace*{-28mm}
\psfig{file=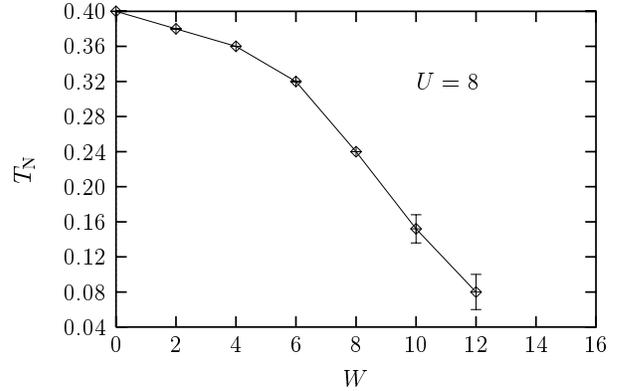,width=135mm,angle=0}
\vspace{-70mm}
\caption{N\'{e}el temperature $T_{\rm N}$ vs. $W$ obtained within the DMFT.}
\end{figure}
\noindent
diagram of which has been
recently studied within the UHF approach.\cite{logan}
A region of metallic AF state in $d=3$ was also
recently identified in the $t-t'$-Hubbard model
without disorder; here it is the NNN hopping amplitude $t'$
that leads to frustration of AF order.\cite{hv}

The robustness of the gapless AF state is also 
seen from a DMFT calculation of the N\'{e}el temperature $T_{\rm N}$,
for a rectangular distribution of $\epsilon_i$, 
as obtained from an extrapolation to zero 
of the inverse AF susceptibility.
Both exact enumeration and Monte Carlo evaluation were employed.
Details of the application of DMFT to the disordered Hubbard model 
have been described earlier,\cite{ulmke}
where binary alloy and semicircular distributions were 
studied.  The variation of $T_{\rm N}$ with
$W$ is shown in Fig. 7 for $U/B=1$.
This corresponds to $U/t=8$ in the two-dimensional case. 
Appreciable spin ordering is evident from the 
fairly high $T_{\rm N}$ even at 
$W \sim U$ where the energy gap vanishes.
Within the DMFT, this closing of the energy gap with disorder,
in the low-temperature ordered state
is seen in Fig. 8 from the single-particle density of states
$N(\omega)$, which was obtained by analytically
continuing the imaginary-time Green's function
using the maximum entropy procedure.\cite{jarrell96}

As expected for weak and intermediate couplings, 
$T_{\rm N}$ is seen to decrease with disorder strength, 
in contrast to the strong-coupling result of 
an enhancement in the magnon energy scale, and hence in $T_{\rm N}$.
Also, the critical disorder strength where $T_{\rm N}$ vanishes 
is seen to be nearly $1.5 \; U$, 
which is intermediate between the critical values of
nearly $2 \; U$ for the milder semicircular distribution
and nearly $U$ for the much more severe binary distribution.\cite{ulmke} 
In fact, for the binary alloy case, the zero-temperature transition is expected 
to occur at $W\approx U$, 
when all sites abruptly become either unoccupied or doubly occupied
and the sublattice magnetization vanishes. 
\begin{figure}
\vspace{-25mm}
\hspace*{-18mm}
\psfig{file=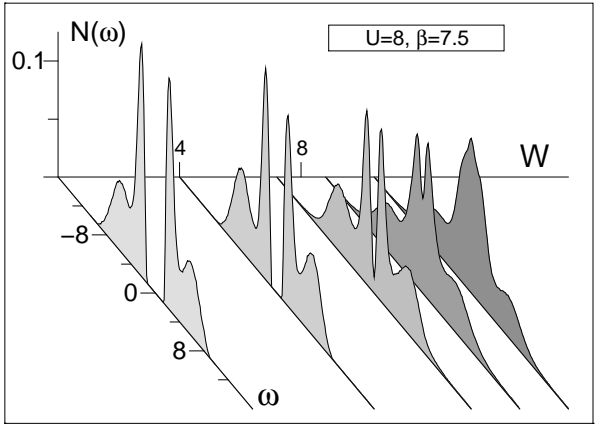,width=105mm,angle=0}
\vspace{-10mm}
\caption{Electronic density of states $N(\omega)$ 
in the low-temperature 
ordered state for several disorder values, 
as obtained within the DMFT. The charge gap closes near $W=U$.}
\end{figure}

\section{Conclusion and Discussion}
Two fundamentally different mechanisms ---
disorder-enhanced delocalization and local-moment quenching ---
were identified to control the magnetic behavior of an AF
in the regimes of weak and strong disorder, respectively.
In the weak disorder regime ($W<U$) 
disorder effects on sublattice magnetization, magnon-mode energies,
and N\'{e}el temperature can be qualitatively understood within
the disorder-enhanced delocalization  effect.
The {\em stabilization} of the N\'{e}el state by disorder in the 
strong correlation limit,
reflected in an {\em enhancement} of the magnon energy scale
and the N\'{e}el temperature, is a striking consequence. 
In this regime the AF state is remarkably robust against disorder, 
particularly in the strong correlation limit.
Low-energy, long-wavelength 
magnon modes are weakly damped, and continue to be well-defined
excitations.

With increasing disorder strength there is a crossover at $W\sim U$, 
characterized by a rapid decrease in sublattice magnetization and 
quenching of local moments due to formation of nonmagnetic sites. 
Driven by band overlap, this local-moment quenching 
can be viewed as the reverse of local-moment formation in 
the disordered metallic state with increasing $U/W$.
These nonmagnetic sites act like spin vacancies
in the antiferromagnet, leading to characteristic spin-dilution behavior.
Vacancy-induced magnon scattering results in 
enhanced transverse spin fluctuations, softened magnon modes,
lowering of N\'{e}el temperature, and strong magnon damping.

As discussed in section VI, 
in the gapless AF state for $W > U$,
the electronic states near the Fermi energy are 
associated with (and localized around) 
the essentially empty and doubly occupied sites.
These states are nonmagnetic in that 
they do not contribute significantly to local moments. 
Therefore it is interesting to note that 
the low-frequency conductivity behavior  
involves these nonmagnetic states near the Fermi energy.
Further investigation of the gapless AF state, 
focussing on the low-frequency conductivity behavior,
magnon damping due to decay into particle-hole excitations,
and an estimation of the critical disorder strength
where AFLRO is destroyed, 
is presently in progress.\cite{next}

\section*{ACKNOWLEDGMENTS}
One of us (A.S.) gratefully acknowledges support 
from the Alexander von Humboldt Foundation through a Research Fellowship.

\section*{Appendix}
\section*{Perturbative analysis in $W/U$: Strong correlation limit} 
In this appendix we treat the disorder term
$V\equiv \sum_{i\sigma}\epsilon_i \hat{n}_{i\sigma}$
in the Hamiltonian 
as a perturbation and obtain disorder corrections  
in the broken-symmetry state up to second order in $W/U$.
With $g^0 _\sigma$ representing the one-electron 
HF Green's function in the pure AF, 
the corresponding Green's function for the disordered AF,
$g_\sigma= g^0 _\sigma/[1-Vg^0 _\sigma]$, 
then yields disorder corrections to electronic densities
and sublattice magnetization.
Similarly, the magnon renormalization within RPA is obtained from 
the correction $\delta \chi^0 \equiv \chi^0_{\rm dis} -
\chi^0_{\rm pure}$ in the zeroth-order antiparallel-spin
particle-hole propagator defined below Eq. (4).

For analytic convenience 
we consider the strong-correlation limit and retain terms of order
$t^2/U^2$ only, at which level the pure AF is
equivalent to the NN Heisenberg model. 
Up to this order the Green's function 
$[g^0_\sigma]_{ij}$ in site-basis 
has only diagonal, nearest-neighbor (NN) and 
next-nearest-neighbor (NNN) matrix elements. 
Only the former two are actually required 
in the strong-coupling analysis, and are given below 
for spin up and site $i$ in the A sublattice.
Expressions for other spin and sublattice cases follow from 
the spin-sublattice symmetry.
These Green's functions are easily obtained by starting with 
the atomic limit, where the bands are at energies $-\Delta$ and $\Delta$,
with $2\Delta=U$, and then obtaining corrections up to second order in the
hopping term $t$.  
The diagonal and NN  matrix elements of the time-ordered
Green's function matrix $[g^0_\sigma]$, 
containing both the advanced (lower band)
and retarded (upper band) parts are,
\begin{eqnarray}
[g_{\uparrow}^0]_{ii}  &=&\frac{1-t^2/\Delta^2}{\omega-(-\Delta) -i\eta}
+\frac{t^2/\Delta^2}{\omega-\Delta +i\eta} \nonumber \\ 
\ [g_{\uparrow}^0]_{i,i+\delta}  &=&\frac{-t/2\Delta}{\omega-(-\Delta) -i\eta}
+\frac{t/2\Delta}{\omega-\Delta +i\eta} .
\end{eqnarray}

We note here that the corrections to one-particle Green's functions 
involve renormalizations of (i) density (wavefunction)
and (ii) energy. Diagrams in which the potential
scattering is interband (involving different Hubbard bands)
result in transfer of spectral weight across the
Hubbard bands, and therefore yield density changes.  
However, the intraband processes (involving all
propagators from the same band) 
represent energy renormalization due to disorder
potential,  and do not contribute to any density change.
In the particle-hole propagator $\chi^0$, while both energy and density
renormalizations need to be considered, diagrams involving only
energy renormalization do not contribute as both the particle and hole
energies are shifted identically by the disorder potential.
\subsubsection{First order}
The first-order correction $g_\sigma^0 V g_\sigma ^0$
yields the following interband contribution 
to the local one-particle Green's function for spin up, 
\begin{eqnarray} 
& & [\delta g^{(1)}_{\uparrow}]_{ii}|_{\rm interband} 
 =2 \left [\frac{1}{\omega-(-\Delta)-i\eta}\
\epsilon_i\ \frac{t^2/\Delta^2}{\omega-\Delta+i\eta} \right .\nonumber \\
 & & +\sum_\delta \left . 
\frac{-t/2\Delta}{\omega-(-\Delta)-i\eta} \ \epsilon_{i+\delta}\ 
\frac{t/2\Delta}{\omega-\Delta+i\eta}\right ],  
\end{eqnarray}
where $i+\delta$ refers to the NN sites of $i$.
The correction to density results from
the spectral weight transferred to the upper band, 
\begin{equation} 
\delta n_{i\uparrow}^{(1)}=\int\frac{d\omega}{2\pi i} e^{i\omega \eta}
[\delta g_{\uparrow}^{(1)}]_{ii}
=-2\frac{t^2}{\Delta^2}\left (\frac{\epsilon_i -\sum_\delta
\epsilon_{i+\delta} /4}{U}\right ) .
\end{equation}

Thus for positive $\epsilon_i$ (and $\epsilon_{i+\delta}=0$, for sake of argument) 
the electron on site $i$ is more
delocalized as its energy is pushed up, leading to the above decrease in 
density. 
A simple way to see this is in terms of the escape probability for the
spin-up electron from site $i$ to its nearest neighbors $i+\delta$ due to the
virtual hopping process.  The net probability of escape changes from
$t^2/U^2$ to $(1/4)\sum_\delta t^2/(U+\epsilon_{i+\delta} -\epsilon_i)^2$, which to
first order in $\epsilon$ precisely yields the above result.  

From the particle-hole symmetry with spin flip in the 
Hubbard antiferromagnet,\cite{uhf1}
the spin-down hole experiences the same effect, except that
the potentials are reversed for the hole. Therefore, there is an
{\em increase} of identical magnitude in the spin-down hole density,
implying an identical decrease in the spin-down particle density.
This leads to a
vanishing change in the local magnetization $m_{\rm HF}$,
and the particle-hole propagator $\chi^0$.  
This cancellation would be absent if the random potential was 
spin-dependent, and is connected to the time-reversal symmetry. 
Thus, 
\begin{equation}
\delta m^{(1)}_{\rm HF}=0 \; ; \;\;\; [\delta \chi^0]^{(1)}=0
\end{equation}
\subsubsection{Second order}
Second-order changes in densities,
$\delta n_{i\sigma}^{(2)}$,
are similarly obtained from the second order correction 
$g_\sigma^0 V g_\sigma ^0 V g_\sigma^0 $. 
As this term is invariant under the transformation $V\rightarrow -V$,
identical changes are obtained for spin-up electron and spin-down hole.
The changes in spin-up and spin-down electronic densities are
therefore equal and opposite, and hence the change in local
magnetization adds up to, 
\begin{equation} 
\delta m_i^{(2)}=
-6\frac{t^2}{\Delta^2}\frac{1}{4}\sum_\delta \frac{(\epsilon_i
-\epsilon_{i+\delta})^2} {U^2}  .
\end{equation}

This result also follows from the second-order 
correction to the net escape probability, discussed earlier
below Eq. (22). 
Configuration averaging, with $\langle 
\epsilon^2\rangle =W^2/12$, 
therefore yields the following quadratic decrease of sublattice magnetization 
with disorder, which is 
suppressed by the small factor $t^2/U^2$.  
As discussed in section III, quantum spin-fluctuation effects do not
substantially modify this result,
which qualitatively agrees with the quantum Monte Carlo 
studies of the disordered Hubbard model.\cite{ulmke2}
\begin{equation} 
\langle \delta m^{(2)}_{\rm HF}\rangle =-4\frac{t^2}{U^2}\frac{W^2}{U^2} .
\end{equation}

Turning now to the second-order corrections to $\chi^0$,
we find that processes containing energy renormalization 
in the fermionic propagators
(either solely, or along with density renormalization) 
cancel, so that the net result to order $(W/U)^2$ is,
\begin{eqnarray} 
& &
[\delta\chi^0]_{ii}^{(2)} = -\frac{t^2}{\Delta^3}\frac{1}{4}\sum_\delta
\frac{(\epsilon_i -\epsilon_{i+\delta})^2} {U^2} , \nonumber \\ 
& &
[\delta\chi^0]_{i,i+\delta}^{(2)} = -\frac{t^2}{\Delta^3}\frac{1}{4}\frac{(\epsilon_i
-\epsilon_{i+\delta})^2} {U^2}  .
\end{eqnarray} 
We note that the 
sum of all NN matrix elements is precisely the diagonal matrix element. 
An immediate consequence of this correlation is that
the Goldstone mode is preserved, as expected from spin-rotational
symmetry, and that generally the effective scattering of low-energy,
long-wavelength magnon modes is weak.
This disorder-induced perturbation $\delta \chi^0$ directly yields
the magnon self energy, and disorder renormalization effects on magnon 
properties are discussed  in sections III and IV.

\end{multicols}

\begin{references}
\bibitem[50]{note1}The coefficient of $x$ was found to be nearly 3 
for long-wavelength magnon modes
within a T-matrix analysis of spin vacancies in the QHAF.\cite{kampf}
\end{references}
\end{document}